\documentclass[twocolumn, showpacs, preprintnumbers, amsmath, amssymb]{revtex4}
\usepackage{graphicx}
\usepackage{dcolumn}
\usepackage{bm}
\usepackage{color}
\begin{document}
\title{Flow in Rough Self-Affine Fractures Joints}

\author{Jos\'e S. Andrade Jr.}
\email{soares@fisica.ufc.br}
\affiliation{Departamento de F\'{\i}sica, Universidade Federal do
Cear\'a, 60451-970 Fortaleza, Cear\'a, Brazil}

\author{Asc\^anio D. Ara\'ujo}
\affiliation{Departamento de F\'{\i}sica, Universidade Federal do
Cear\'a, 60451-970 Fortaleza, Cear\'a, Brazil}

\author{Fernando A. Oliveira}
\affiliation{International Center of Condensed Matter Physics and 
Instituto de F\'{\i}sica, Universidade de Bras\'{\i}lia, CP 04667,
70919-970 Bras\'{\i}lia DF, Brazil}

\author{Alex Hansen}
\affiliation{Department of Physics, Norwegian University of Science and
Technology, N--7491 Trondheim, Norway}

\date{\today}
\begin{abstract}
We investigate viscous and non-viscous flow in two-dimensional
self-affine fracture joints through direct numerical simulations of
the Navier-Stokes equations. As a novel hydrodynamic feature of this
flow system, we find that the effective permeability at higher
Reynolds number to cubic order, falls into two regimes as a function
of the Hurst exponent $h$ characterizing the fracture joints. For
$h>1/2$, we find a weak dependency whereas for $h<1/2$, the dependency
is strong. A similar behavior is found for the higher order
coefficients. We also study the velocity fluctuations in space of a
passive scalar. These are strongly correlated on smaller length
scales, but decorrelates on larger scales. Moreover, the fluctuations
on larger scale are insensitive to the value of the Reynolds number.
\end{abstract} 
\maketitle
The transport of oil in carbonate reservoirs is dominated by flow
through internal fractures. As about 10-15\% of the world's known
reservoirs consist of carbonates and that these contain about 50\% of
the remaining oil, it is surprising that this problem has received so
little attention as it has from not only the physics community, but
also the engineering communities. It is not uniquely from a practical,
economical point of view that this is a worthwhile problem to
study. The progress made in fracture morphology over the last years,
has uncovered a large number of basic questions to be answered, and
the present problem is one of these. It has become increasingly clear
that the morphology of brittle fractures follows certain scaling laws
first observed in the mid-eighties \cite{Man84}, demonstrating that
such surfaces are {\it self-affine.\/} That is, by rescaling the in-plane
length scale $x$ by $\lambda x$ requires the out-of-plane scale $y$ to
be rescaled by $\lambda^h y$ for the fracture surface to remain
statistically unchanged, with $h$ being the roughness exponent. In
the early nineties, it was suggested that $h$ is universal and has
a value of 0.8 \cite{Bou90}. There is today some evidence that 
other values of the Hurst exponent may occur depending on the material 
and how the fracture has been produced \cite{PBB06}, thus suggesting 
more than one universality class.

\begin{figure}[b]
\includegraphics[width=7cm]{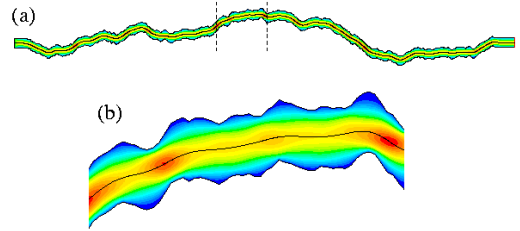}
\caption{(a) Contour plot of the local velocity magnitude along
a typical realization of the self-affine rough channel ($h=0.8$)
subjected to low Reynolds conditions. Fluid is pushed from left to
right. The colors ranging from blue to red correspond to low and high
velocity magnitudes, respectively. The solid line in the center of the
channel corresponds to the streamline that divides the flux into two
equal parts. The zoom shown in (b) reveals the presence of high
velocity spots in the flow field induced by the occurrence of high
slopes in the channel geometry.}
\label{fig1}
\end{figure}

For single-phase fluid flow in porous media and fractures, it is
common to characterize the system in terms of Darcy's law
\cite{Dul79,Sah94}, which assumes that a {\it global\/} index,
the permeability $k$, relates the average fluid velocity $V$ with the
pressure drop $\Delta P$ across the system, $V = -k \Delta P/\mu L$.
Here $L$ is the length of the sample in the flow direction and $\mu$
is the viscosity of the fluid. However, in order to understand the
interplay between porous structure and fluid flow, it is necessary to
examine {\it local} aspects of the pore space morphology and relate
them to the relevant mechanisms of momentum transfer (viscous and
inertial forces). Flow in self-affine faults was to our knowledge
first discussed by Roux {\it et al.\/} \cite{Rou93}, where they
considered a fracture fault consisting of two matching walls that
have been moved along the fracture plane with respect to each
other. If the in-plane movement is $x$, then -- due to the
self-affinity of the fault -- the amplitude of the fault will be
proportional to $x^h$. It was claimed in \cite{Rou93} that the
permeability of a two-dimensional fracture fault would scale as
$x^{2h}$, based on using the self-affinity to determine the scaling
properties of an effective channel width. However, numerical studies
by Gutfraind {\it et al.\/} \cite{Gut95} using lattice gas methods,
demonstrated that the permeability should be controlled by the
narrowest neck in the system. In a subsequent work \cite{Dra00}, the
permeability of self-affine rough fractures with wide and narrow
apertures has been investigated analytically and confirmed through
numerical simulations with the lattice Boltzmann method. The flow in
self-affine fracture {\it joints\/,} where the two matching fractures
are moved apart by a distance $w$ but without any relative movement in
fracture plane ($x=0$), has been previously studied by Skjetne {\it et
al.\/} \cite{Skj99}.

\begin{figure}[t]
\includegraphics[width=6cm]{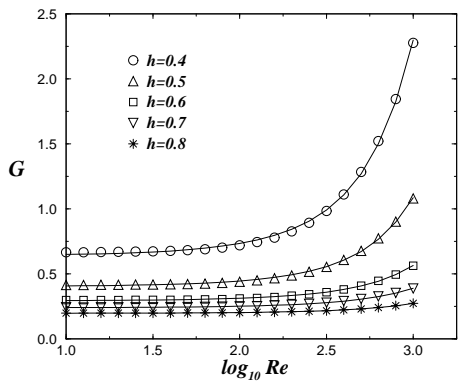}
\caption{Dependence of the hydraulic resistance $G \equiv -\Delta P 
w^{2}/\mu V L$ on the Reynolds number for different values of the
roughness exponent $h$. In all cases, the plateau corresponding to
Darcy's law (constant $G$) is followed by a nonlinear regime that
reflects the effect of convection on the flow. The error bars are
smaller than the symbols and the solid lines are the best fit to the
data of Eq.~(\ref{Eq_conductance}).}
\label{fig2}
\end{figure}

Strictly speaking, the concept of permeability as a global index for
flow in fractures should be restricted to viscous flow (linear)
conditions. More precisely, Darcy's law should only be applicable for
flow at sufficiently low Reynolds number, defined here as
$Re\equiv{\rho V w / \mu}$, where $w$ is the fracture opening. It is
well known, however, that the role of inertial forces (convection) to
flow in disordered media should be examined in the framework of the
laminar flow regime, before assuming that fully developed turbulence
effects are already present \cite{Dul79,Koc97,And99}. Much less effort
has been dedicated to address this problem in the specific context of
flow through rough fractures. For instance, the computational fluid
dynamics simulations presented in Ref.~\cite{Skj99} indicate that
vanishing weak and strong inertial flows can be quantitatively
described in different ways, namely the Darcy, weak inertia and
Forchheimer empirical equations, respectively. Here we investigate by
direct simulation of the Navier-Stokes equations the departure from
Darcy's law in laminar flow through self-affine fracture joints. We
confirm that the physical description underlying a classical cubic
equation provides a legitimate correlation for the flow in the
fracture over a wide range of Reynolds number conditions. We then
demonstrate that it is also possible to characterize this transition
from linear to nonlinear behavior in terms of the spatial correlations
in the fluid velocity. This allows us to elucidate certain
characteristics of the fluid flow phenomenon in fracture fault
geometries that have not been studied before.

The self-affine surfaces are generated here with a Fourier method
\cite{Vos85} for which we specify the length $N$ and width $m$, both in 
terms of the number of nodes, and the roughness exponent $h$. The
surface is given by $y_i$ (measured in units of the lattice constant),
and $m=\max y_i - \min y_i$. Since the lattice constant is $\delta$, the
length of the system is $L=N\delta$ and its amplitude is given by 
$a = m\delta$. When $L$ is to be changed while $\delta$ is to
be kept fixed, we scale $N\to\lambda N$ and $m\to\lambda m$. On the
other hand, when $\delta$ is to be changed, while $L$ is to be kept
fixed, we let $N\to\lambda N$ and keep $m$ fixed. The fracture opening
$w$ is kept fixed. 

\begin{figure}
\includegraphics[width=6cm]{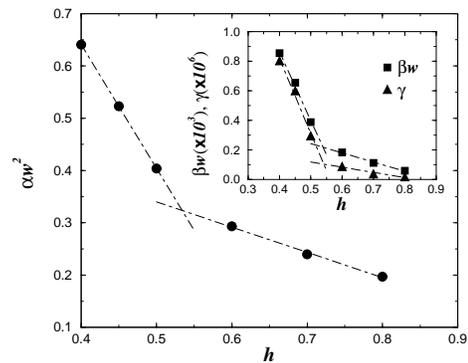}
\caption{(a) Variation of the parameter $\alpha$ as in  
Eq.~(\ref{Eq_conductance} with the exponent $h$ characterizing the
roughness of the channel geometry. The two dashed lines correspond to
different regimes dictated by the nature of the correlations. The
crossover at $h \approx 0.5$ delimits anti-correlated ($h<0.5$) and
correlated ($h>0.5$) signals. The inset shows that the parameters
$\beta$ and $\gamma$ of Eq.~(\ref{Eq_conductance}) behave in a very 
similar way.}
\label{fig3}
\end{figure}

The mathematical description for the fluid mechanics in the
interstitial porous space is based on the assumptions that we have a
continuum, Newtonian and incompressible fluid flowing under steady
state conditions. Thus, the Navier-Stokes and continuity equations
reduce to
\begin{equation}
\rho~{\bf{u\cdot\nabla u}} = -{\nabla p} + \mu~{\bf{\nabla}}^{2}{\bf{u}}~,
\label{Eq_momentum}
\end{equation}
\begin{equation}
{\bf{\nabla\cdot u}}=0~,
\label{Eq_continuity}
\end{equation}
where {\bf u} and $p$ are the local velocity and pressure fields,
respectively, and $\rho$ is the density of the fluid. In our
simulations, we consider non-slip boundary conditions along the entire
solid-fluid interface. In addition, the changes in velocity rates are
assumed to be zero at the exit $x=L$ (gradientless boundary
conditions), whereas a uniform velocity profile, $u_{x}(0,y)=V$ and
$u_{y}(0,y)=0$, is imposed at the inlet of the channel.

The numerical solution of Eqs.~(\ref{Eq_momentum}) and
(\ref{Eq_continuity}) for the velocity and pressure fields in the pore
space is obtained through discretization by means of the control
volume finite-difference technique \cite{Pat80}. Considering the
complex geometries involved, we build an unstructured mesh of
triangular grid elements based on a Delaunay network. For the largest
fracture investigated, namely $L=512$, around $10^6$ cells are
necessary to generate satisfactory results when compared with
numerical meshes of larger resolution. The convergence criteria used
in the simulations are defined in terms of residuals, i.e., the degree
up to which the conservation equations are satisfied throughout the
flow field. In all our simulations, convergence is considered to be
achieved only when each of the residuals falls below $10^{-6}$.

In Fig.~\ref{fig1} we show the contour plot of the velocity magnitude
in a typical self-affine channel under viscous flow conditions, i.e.,
at very low Reynolds number ($Re=10$). Clearly, the spots of high
velocity correspond to those regions with high slopes in the channel
due to their reduced effective areas for flow, namely the
cross-sections orthogonal to the walls. Also shown in Fig.~\ref{fig1}
is the streamline that divides the flow exactly into two zones of
equal flux. As discussed later, this line will be adopted here as a
reference location to study the spatial correlations in the local
fluid velocity for different Reynolds conditions.

The approach we adopt here to macroscopically characterize the effect
of convection on flow through the self-affine channel is to employ the
cubic relation
\begin{equation}
-\frac{\Delta P}{L} = \alpha \mu V + \beta \rho {V^2} + 
\frac{\gamma \rho^{2} V^{3}}{\mu}~, 
\label{Eq_cubic}
\end{equation}
where the coefficient $\alpha$ corresponds to the reciprocal of the
permeability of the porous material and the two remaining terms
containing $\beta$ and $\gamma$ can be interpreted, respectively, as
second and third order corrections that should account for the
contribution of inertial forces in the fluid flow. At sufficiently low
Reynolds, Eq.~(\ref{Eq_cubic}) reduces to Darcy's law. Rewriting
(\ref{Eq_cubic}) in terms of the Reynolds number we obtain,
\begin{equation}
G = \alpha w^{2} + \beta w Re + \gamma Re^{2}~,
\label{Eq_conductance}
\end{equation}
with $G \equiv -\Delta P w^{2}/\mu V L$ being a dimensionless measure
of the {\it hydraulic resistance}. Fig.~\ref{fig2} shows the results
of our flow simulations in terms of the variables $G$ and $Re$ for
different values of the roughness exponent $h$. After computing and
averaging $G$ over a total of 10 realizations for each value of $h$
and a wide range of Reynolds numbers, we fit the results with
Eq.~(\ref{Eq_conductance}) to estimate the coefficients $\alpha$,
$\beta$ and $\gamma$. In agreement with real flow experiments, we
observe a transition from constant $G$ (Darcy's law) to nonlinear flow
behavior at a value of $Re$ that depends significantly on the
roughness. As shown in Fig.~\ref{fig3}, the coefficient $\alpha$
generally decreases with the roughness exponent $h$. There is,
however, a clear crossover at a value of $h \approx 0.5$ separating
two distinct regimes that characterize the influence of the channel
geometry on the parameter $\alpha$. Interestingly, the value $h=0.5$
corresponds exactly to the transition point between anti-correlated
($h<0.5$) and correlated ($h>0.5$) self-affine interfaces
\cite{Hav96}. To the best of our knowledge, this is the first time
that a clear connection can be drawn in order to relate the degree of
correlation in the interface geometry and the single-phase flow
behavior through self-affine fractures. The results shown in the inset
of Fig.~\ref{fig3} reveal that the coefficients $\beta$ and $\gamma$
also decrease with $h$ and display similar crossovers between
correlated and anti-correlated geometries.

\begin{figure}[t]
\includegraphics[width=7cm]{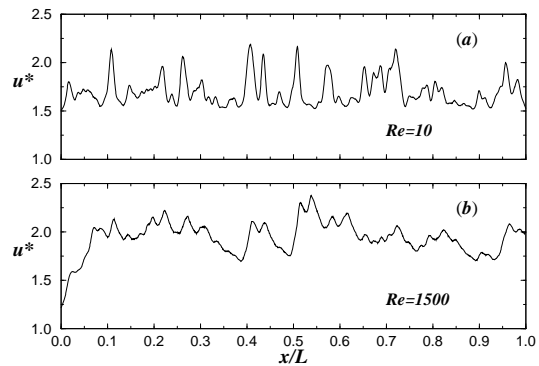}
\caption{(a) Profile of the velocity magnitude along the streamline
located at the center of the rough channel (see Fig.~\ref{fig1}) for a flow
field calculated at low Reynolds conditions ($Re=10$). (b) The same 
as in (a), but for $Re=1500$.}
\label{fig4}
\end{figure}

Next we study the fluctuations in velocity of a massless particle
(i.e., a passive tracer) released right at the center of the channel
inlet. As mentioned earlier, this particle will then follow a
trajectory that coincides with the streamline dividing the flow into
two regions of equal flux (see Fig.~\ref{fig1}). In Fig.~\ref{fig4},
we show the variation of its normalized velocity magnitude
$u^{*}=u/V$, along the main flow direction $x$ in a typical
realization of the rough channel and for two different values of the
Reynolds number. At low $Re$ (Fig.~\ref{fig4}a), the location and
intensity of the velocity peaks essentially correspond to the spatial
variation in amplitude of the local slopes along the fracture. At high
$Re$ (Fig.~\ref{fig4}b), the situation becomes quite different. Due
to inertia, the effect on the flow field of the local channel geometry
reveals a {\it persistent} behavior in the local velocity
fluctuations, when compared to the results obtained at low $Re$ 
(Fig.~\ref{fig4}a). More precisely, whenever a sudden increase in velocity
is observed due to the presence of a narrow constriction in the
channel, the signal tends to decay more slowly at higher $Re$
conditions, before the particle experiences another substantial
amplitude fluctuation. It is interesting to note that the same
sequence of peaks and valleys presented in Fig.~\ref{fig4}a can also
be observed in Fig.~\ref{fig4}b, but with the difference that the
background velocity is generally much higher at high $Re$ values
because of inertial effects.

\begin{figure}[t]
\includegraphics[width=6cm]{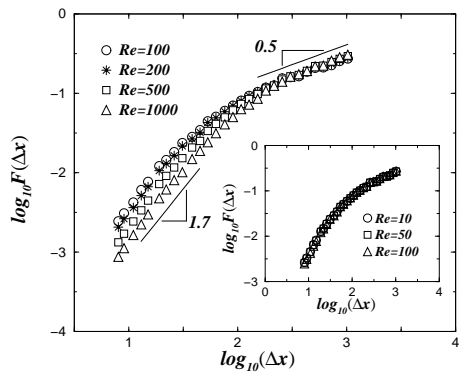}
\caption{(a) Logarithmic plot of the detrended fluctuation function 
$F(\Delta x)$ calculated from 10 realizations of velocity magnitude
profiles (see Fig.~\ref{fig4}) and four different values of the
Reynolds number. The two straight lines show the best power law fits
to the data in the scaling regions. The difference in the slopes
indicates the passage from a highly correlated ($\zeta \approx 1.7$)
to an uncorrelated ($\zeta \approx 0.5$) series. The inset shows that
the function $F(\Delta x)$ is invariant with $Re$ for $Re<100$.}
\label{fig5}
\end{figure} 

Due to the interplay between flow and the self-affine characteristic
of the fracture interfaces utilized here \cite{Hav96}, one should
expect the velocity profiles shown in Fig.~\ref{fig4} to contain a
certain degree of correlation. More precisely, long-range power-law
correlations in velocity magnitude are identified and quantified here
by means of the {\it detrended fluctuation analysis} (DFA)
\cite{Pen94}. According to this method, one can avoid the spurious
observation of long-range correlations due to non-stationarities by
integrating the time series and mapping it to a self-affine stochastic
process. The velocity profile $u^{*}(x)$ is integrated, after
subtracting the overall velocity average, and divided into boxes of
equal length $\Delta x$. The local linear trend is then calculated as
the least-squares straight line that fits the data within each
interval. The difference between the integrated time series and the
local trend in each box gives the detrended time series, from which we
can compute the root-mean-square fluctuation $F(\Delta x)$ \cite{Pen94}. 
Scaling is present if $F(\Delta x)$ has a power-law dependence on 
$\Delta x$, $F(\Delta x) \simeq (\Delta x)^{\zeta}$, where the exponent 
$\zeta$ characterizes the nature of the long-range correlations.

The results in Fig.~\ref{fig5} show that, regardless of the Reynolds number,
the function $F(\Delta x)$ displays a highly correlated power-law
regime ($\zeta=1.7 \pm 0.03$) at small length scales followed by a
typically uncorrelated scaling ($\zeta=0.5 \pm 0.01$) at larger
values of the $\Delta x$ window. The only difference is that, above a
sufficiently high value of $Re$ (see the inset of Fig.~\ref{fig5}), the
crossover from correlated to uncorrelated behavior starts to increase
with $Re$. This is compatible with our previous qualitative analysis
based on the simple inspection of the profiles in Fig.~\ref{fig4}. 
The first (highly correlated) regime is a direct consequence of the channel
self-affinity inducing long-range correlations in tracer velocity,
where a large exponent denotes the fractional-Brownian aspect of the
signal.

In summary, we have presented a numerical study of single-phase flow
in self-affine fracture joints. We find that each of the coefficients
$\alpha$, $\beta$ and $\gamma$, characterizing the cubic
generalization of the Darcy equation (\ref{Eq_cubic}) depends nearly
linearly on the Hurst exponent $h$ in two distinct regimes that meet
at $h \approx 0.5$. We also follow a passive scalar particle released
at the center of the channel, finding that its velocity fluctuations
are characterized by strong spatial correlations on small scales,
crossing over to uncorrelated random-walk-like behavior on larger
scale. The large-scale behavior is insensitive to $Re$.

We thank Josu\'{e} Mendes Filho and Andr\'e Moreira for discussions,
the Brazilian agencies CNPq (CT-PETRO/CNPq), CAPES, FINEP and FUNCAP,
and the Norwegian agency NFR for financial support.



\begin{thebibliography}{99}

\bibitem{Man84} 
B. B. Mandelbrot, D. E. Passoja, and A. J. Paullay, Nature {\bf
308}, 721 (1984); S. R. Brown and C. H. Scholz, J. Geophys. Res. {\bf
90}, 12575 (1985).

\bibitem{Bou90} 
E. Bouchaud, G. Lapasset, and J. Plan{\'e}s, Europhys. Lett. {\bf 13},
73 (1990); K. J. M{\aa}l{\o}y, A. Hansen, E. L. Hinrichsen, and
S. Roux, Phys. Rev. Lett. {\bf 68}, 213 (1992); J. Schmittbuhl,
S. Gentier, and S. Roux, Geophys. Res. Lett. {\bf 20}, 639 (1993);
B. L. Cox and J. S. Y. Wang, Fractals {\bf 1}, 87 (1993);
J. Schmittbuhl, F. Schmitt, and C. H. Scholz, J. Geophys. Res. {\bf
100}, 5953 (1995).

\bibitem{PBB06} L.\ Ponson, D.\ Bonamy and E.\ Bouchaud, Phys.\ Rev.\
Lett.\ {\bf 96}, 035506 (2006).

\bibitem{Dul79} 
F. A. L. Dullien, {\it Porous Media - Fluid Transport and Pore
Structure} (Academic, New York, 1979).

\bibitem{Sah94}
M. Sahimi, {\it Applications of Percolation Theory\/} (Taylor \&
Francis, London, 1994); M. Sahimi, {\it Flow and Transport in Porous
Media and Fractured Rock\/} (VCH, Boston, 1995).

\bibitem{Rou93} 
S. Roux, J. Schmittbuhl, J.-P. Vilotte and A. Hansen, Europhys. Lett.
{\bf 23}, 277 (1993).

\bibitem{Gut95} 
R. Gutfraind and A. Hansen, Transp. Porous Media {\bf 18}, 131
(1995); R. Gutfraind, I. Ippolito and A. Hansen, Phys. Fluids {\bf
7}, 1938 (1995).

\bibitem{Dra00} G. Drazer and J. Koplik, Phys. Rev. E {\bf 62}, 
8076 (2000).

\bibitem{Skj99} E. Skjetne, A. Hansen and J. S. Gudmundsson,
J. Fluid Mech. {\bf 383}, 1 (1999).

\bibitem{Koc97} D. R. Koch and A. J. C. Ladd, J. Fluid Mech. {\bf 349},
31 (1997).

\bibitem{And99}
J. S. Andrade Jr., U. M. S. Costa, M. P. Almeida, H. A. Makse and
H. E. Stanley, Phys. Rev. Lett. {\bf 82}, 5249 (1999).

\bibitem{Vos85} 
R. F. Voss in R. A. Earshaw, Ed. {\it Fundamental Algorithms for
Computer Graphics\/} (Springer Verlag, Berlin, 1985).

\bibitem{Pat80}
S. V. Patankar, \textit{Numerical Heat Transfer and Fluid Flow} 
(Hemisphere, Washington DC, 1980); The {FLUENT} (trademark of
{FLUENT} Inc.) fluid dynamics analysis package has been used in
this study.

\bibitem{Hav96}
\textit{Fractals and Disordered Systems} 2nd ed., edited by
A. Bunde and S. Havlin (Springer-Verlag, New York, 1996).

\bibitem{Pen94}
C.-K. Peng, S. V. Buldyrev, S. Havlin, M. Simons, H. E. Stanley, 
and A. L. Goldberger, Phys. Rev. E {\bf 49}, 1685 (1994).


\end{thebibliography}
\end{document}